\documentclass[fleqn,12pt]{wlscirep}
\usepackage{color}
\title{The foundations of statistical physics: entropy, irreversibility, and inference}

\author[1,2]{Jonathan Pachter}
\author[1]{Ying-Jen Yang}
\author[1,2,*]{Ken Dill}
\affil[1]{Laufer Center for Physical and Quantitative Biology, Stony Brook University, NY, USA}
\affil[2]{Department of Physics and Astronomy, Stony Brook University, NY, USA}
\affil[*]{e-mail: dill@laufercenter.org}

\begin{abstract}
Statistical physics aims to describe properties of macroscale systems in terms of distributions of their microscale agents.  Its central tool is the maximization of entropy, a variational principle.  We review the history of this principle, first considered as a law of nature, more recently as a procedure for inference in model-making.  And while equilibria (EQ) have long been grounded in the principle of Maximum Entropy (MaxEnt), until recently no equally foundational generative    principle has been known for non-equilibria (NEQ).  We review evidence that the variational principle for NEQ is Maximum Caliber.  It entails maximizing \textit{path entropies}, not \textit{state entropies}.  We also describe the role of entropy in characterizing irreversibility, and describe the relationship between MaxCal and other prominent approaches to NEQ physics, including Stochastic Thermodynamics (ST), Large Deviations Theory (LDT), Macroscopic Fluctuation Theory (MFT), and non-extensive entropies.
\end{abstract}
\begin{document}

\maketitle

\section*{Introduction}
Statistical physics emerged in the mid-1800's as a way to explain and predict macroscopic material phenomena from the physics of their constituent atoms, molecules, or other particles.  Initially, it focused primarily on states of equilibrium (EQ) and utilized the principle of Maximum Entropy (MaxEnt), which is equivalent to some formulations of the Second Law of Thermodynamics.  Its methods and practice are well-known\cite{huang_statistical_1987,hill_statistical_1987, mcquarrie_statistical_2000,dill_molecular_2010,pathria_statistical_2021}.  But two foundational questions have been subject to changing answers over time.  First, what is the reason for using the Boltzmann-Gibbs (BG) form of entropy in statistical physics, and why is it maximized?  Is it a law of physics or a law of inference?  Second, what general procedure can predict the behaviors of non-equilibrium (NEQ) systems?  Driving an interest in these foundational matters is increasing applications of MaxEnt to biology\cite{Martino_Martino_2018}, ecology\cite{Harte_Newman_2014}, economics\cite{Golan_Harte_2022}, active matter\cite{fodor_irreversibility_2022}, and artificial intelligence and neural networks\cite{tsai_path_2022}.  In contrast to previous reviews of the mechanics and applications of EQ (Maximum Entropy) and NEQ (Maximum Caliber)\cite{presse_principles_2013,ghosh_maximum_2020}, here we review the history of their foundations and frame the context of related sub-fields.  \\

We first note that two disparate perspectives are used in statistical physics.  From its earliest days, it was framed in terms of Hamilton's equations of the positions and momenta of collisional particles, their conservation of energy through the Liouville equation, and puzzles of irreversibility and disorder\cite{bagchi2018statistical, reichl2016modern, doi:10.1143/JPSJ.12.570}. The second perspective -- which is our main focus here -- is the language of probabilities, which is the basis for most present-day problem-solving.  The probabilistic basis extends the reach of statistical physics to problems and processes well beyond collisional particles as functions of temperature and pressure.  Newtonian mechanics is an unnecessarily limiting starting point for developing the most general principles of model-making, especially for many-body living systems whose constituents are themselves high-dimensional\cite{yang_statistical_2023}.  Furthermore, the probability framework gives tools that are practical and simple to apply without sacrificing rigor.  \\

\section*{A Tale of Two Entropies: The Foundations of Equilibrium Statistical Physics}
At the heart of statistical physics is a remarkable equality between two seemingly unrelated quantities -- a macroscopic physical property of matter and a mathematical property of probability distributions -- which are both now called \textit{entropy:}
\begin{equation}
    \Delta S_\mathrm{Clausius}=\Delta S_\mathrm{BG}^*
    \label{eq: two entropies}
\end{equation}
On the left side of Eq. \eqref{eq: two entropies} is the change in Clausius entropy, a macroscopic physical quantity defined by Rudolph Clausius in 1855, encapsulating certain transformations between thermodynamic equilibria.  On the right side is the change in maximized Boltzmann-Gibbs (BG) entropy, a mathematical property of a probability distribution over microscopic degrees of freedom of a system.  In the following sections, we define these two entropies and review the history and basis for the principle embodied in Eq. \eqref{eq: two entropies}. \\

\paragraph{The \textit{Clausius Entropy} describes macroscopic processes.}  At the foundation of thermodynamics is an inequality due to Clausius, accompanied by a definition of \textit{reversible, quasi-equilibrium processes} as a very special limiting case of dynamics.  Clausius' theorem states that the integral of the infinitesimal heat a system dissipates to its environment, divided by temperature, is always non-negative along a cyclic transformation, and is only zero for a reversible cyclic process -- a result that relates to both Kelvin's and Carnot's more qualitative statements of the Second Law of Thermodynamics\cite{huang_statistical_1987}:
\begin{equation}
    \label{Clausius theorem}
    \oint_{\mathrm{general}} \frac{\delta Q}{T} \ge \oint_{\mathrm{reversible}} \frac{\delta Q}{T} = 0.
\end{equation}
This result can be used to define the change in Clausius (or thermodynamic) entropy as this heat integral along any reversible trajectory, which must be less than or equal to the same integral for a general trajectory connecting the same two thermal equilibrium states, due to the theorem above:
\begin{equation} 
-\Delta S_{\mathrm{Clausius}} = \int_{\mathrm{reversible}} \frac{\delta Q}{T} \le \int_{\mathrm{general}} \frac{\delta Q}{T}.
\label{delta S_C lower bounds heat dissipation}
\end{equation}
While this property of $\Delta S_{\mathrm{Clausius}}$ can be rearranged to explain heat flow from hotter to colder systems, it contains no information about time dependence, such as how long two bodies at different temperatures would take to equilibrate with each other, and it says nothing about the underlying microscopic constituents of matter. \\

\paragraph{Statistical physics arose from the need for probability distributions.}
The need for probability distributions in connecting macroscale to microscale behavior was already evident in the 1850's when James Clerk Maxwell developed the \textit{Kinetic Theory of Gases}.  Whereas the \textit{mean velocity} of particles in a gas at equilibrium is zero (and therefore uninformative), the \textit{variance} of velocity was found to be related to the temperature $T$ through $\langle v^2 \rangle = 3kT/2m$ for an ideal gas.  Where there is a variance, there is uncertainty, and thus a need for distributions over microscopic agents to explain some macroscopic behaviors.  \\

Throughout its history, statistical physics has been enmeshed within a broader controversy about the nature of probabilities themselves. 
Probabilities are mathematically defined as any fractional quantities, associated with events, that can be combined into other Boolean collections of independent and mutually exclusive events through the addition and multiplication logical rules of probability\cite{kolmogorov_foundations_2018}. The controversy is in whether probabilities can only be descriptions of \textit{frequencies of replicatable events,} as when a dice is rolled multiple times, or can more broadly also describe \textit{inferences about non-repeatable events}, i.e. \emph{frequentist} versus \emph{Bayesian}.  While the \textit{methodology} of statistical physics has been stable since the earliest days, how that methodology is justified on principle has changed -- initially primarily as frequentist, with the Bayesian inferential gaining more traction over the decades.  \\

\section*{Gibbs' ensembles: Probabilities interpreted as frequencies}

Following preliminary work by L. E. Boltzmann, J. C. Maxwell, and H.W. Watson\cite{Inaba_2015}, J. Willard Gibbs constructed a framework of ``ensembles" to encapsulate statistical physics theory, which employed frequencies, the prevailing language for probabilities at the time. The concept of an ensemble is simple.  Consider a system with $M$ possible microstates, e.g. $M$ possible ways of arranging its microscopic constituents. Now, imagine we have $N$ identical copies of our system, each subject to the exact same macroscopic/thermodynamic external conditions.  We can count the number of times $n_m$ we see microstate $m$ within this ensemble of copies, giving us a set of occurrence numbers for all the different $m$. The number of different ways we could observe the same set of occurrences $\{n_m\ |\ m=1,2,...,M\}$ with $\sum_m n_m =N$ is a multinomial:
\begin{equation} \label{Multi}
W = {N!} \bigg/ \left(\prod\limits_{m=1}^{M} n_m !\right).
\end{equation}
Defining the frequency/proportion of each outcome as $f_m = n_m/N$, we can calculate the original Boltzmann definition of entropy in terms of multiplicity $S = \ln W / N$.
Taking the limit of infinitely large ensemble and using Stirling's approximation for factorials in the limit $N \rightarrow \infty$, we get the \textit{Boltzmann-Gibbs} (BG) entropy:
\begin{equation}
S_\mathrm{BG} = \lim_{N \rightarrow \infty} \frac{\ln W}{N} = - \sum\limits_{m=1}^{M} f_m \ln f_m.
\label{eq:fs}
\end{equation}
In the standard treatment, the Boltzmann distribution for a system in a heat bath is derived by maximizing the BG entropy. As detailed below in our discussion on large deviation theory, the set of frequencies that maximize $S_\mathrm{BG}$ corresponds to what is most likely to be observed (see below), subject to constraints on the average energy and normalization of the frequencies, which are imposed using Lagrange multipliers:
\begin{equation}
S_\mathrm{BG} =  -\sum_{m=1}^M f_m \ln f_m - \beta \left (\sum_{j} f_m \varepsilon_m - \langle \varepsilon \rangle \right) + \alpha  \left( \sum_{m=1}^M f_m - 1 \right).
\label{eq:maxent}
\end{equation}
The maximized value $S_\mathrm{BG}^*$ is achieved for the Boltzmann distribution:
\begin{equation}
    f_m^* = \frac{e^{-\beta \varepsilon_m}}{Z} 
    \label{eq: star}
\end{equation}
where
\begin{equation}
    Z = \sum_{m=1}^M e^{-\beta \varepsilon_m}.
    \label{eq:Z}
\end{equation} \\

\paragraph{Large Deviation Theory and related problems of data inference}

It is useful here to introduce \textit{Large Deviations Theory} (LDT).  One reason is that while means and variances often summarize key aspects of probability distributions, the rare events found in the tails of distributions are often of great interest as well. While LDT has applications in statistical physics \cite{touchette_large_2009}, it is particularly relevant in what could be called ``data-rich'' situations, in which events are replicable and  Gibbs' ensemble approach is imitated in real life, leading to an ``idealized infinite dataset'' \cite{lu_emergence_2022,yang_statistical_2023}.  \\

Another usage of LDT is to make the connection, noted above, between maximum multiplicity and most probable observation.  Given a prior distribution $\{q_m\}$, which in textbook treatment was chosen to be uniform according to the \emph{principle of equal a priori probabilities}\cite{pathria_statistical_2021} , the probability of observing a particular set of frequencies $\{f_m\}$ in the Gibbs ensemble with respect to the prior has the following asymptotic form\cite{sanov_probability_1958}:
\begin{equation} \label{LDT expression of P(f)}
    \mathrm{Prob}(f) = \exp \left[-N \cdot D(f||q) + o(N)\right] = \exp\left[ N \cdot S_{\mathrm{BG}}(f||q) + o(N)\right].
\end{equation}
where $o(N)$ are terms that are much smaller than $N$ in the limit of large $N$.
When $N \rightarrow \infty$, this probability distribution is exponentially dominated by the frequency $f$ that is closest to the prior $q$, as measured by the KL divergence.  One can perform a quadratic approximation of $S_\mathrm{BG}$ near $q$ for \textit{small, linear deviations}, leading to a Gaussian probability\cite{landau_statistical_1980}, as dictated by the central limit theorem.  But Eq. \eqref{LDT expression of P(f)} contains information for $f$ further away from $q$, hence the name \textit{large deviations}.  When newly acquired data constrains the possible frequency $f$, the most-probable frequency $f^*$ under the constraints is the one that uniquely maximizes the entropy $S_{\mathrm{BG}}(f||q)$. 
In such a ``data-rich" situation, entropy maximization in Gibbs' framework is equivalent to Bayesian model updating\cite{lu_emergence_2022,yang_statistical_2023}.  This LDT formulation is useful in generalizing Gibbs' framework to infer Markov processes from correlated data \cite{csiszar_conditional_1987,chetrite_nonequilibrium_2013,yang_statistical_2023}. \\

\paragraph{The equality of the two entropies.}  We return to Eq. \eqref{eq: two entropies}, the equivalence between two entropies.  Substituting the posterior $f_m^*$ of Eq. \eqref{eq: star} into Eq. \eqref{eq:maxent} for the BG entropy yields the maximized entropy $S_{\mathrm{BG}}^* = \beta \langle \varepsilon \rangle + \ln Z$.  Consider a process in which we transform the system by slightly perturbing the energy levels by $\delta \varepsilon_j$.  We can compute the change in $\ln Z$ as
\begin{equation}
\delta \ln Z = \sum_{m=1}^M \frac{1}{Z} \frac{\partial Z}{\partial \epsilon_m} \ \delta \varepsilon_m = - \beta \sum_{m=1}^M \frac{1}{Z} q_m e^{-\beta \varepsilon_m} \ \delta \varepsilon_m = - \beta \sum_{m=1}^M f_m^* \ \delta \varepsilon_m = -\beta \delta W
\end{equation}
where $\delta W$ is the work performed on the system in a reversible process.  Thus the change in maximized BG entropy is
\begin{equation}
\delta S_{\mathrm{BG}}^* = \beta \delta \langle \varepsilon \rangle + \delta \ln Z = \beta \delta(\langle \varepsilon \rangle - W) = -\beta \delta Q = \delta S_{\mathrm{Clausius}}
\end{equation}
where $\delta Q$ is the heat dissipated in this reversible process, and we equate the Lagrange multiplier $\beta$ with the inverse temperature $1/(k_B T).$ \\

The ensemble approach has been useful for predicting macroscopic properties of microscopic models. However, its conceptualization of probabilities as frequencies limits its applicability beyond replicable events. The ensemble approach envisions a set of infinite copies of a system, but in reality we often have only one system.  The replicates could arise from different snapshots of a system spaced out in time.  This conception, however, requires two companion assumptions which restrict its generality: first, the \textit{ergodic hypothesis}, which equates time averages with ensemble averages, and second, the independence of replicates due to chaotic interactions or random collisions.  These limitations can be circumvented by framing statistical physics instead as a matter of drawing inferences in probabilistic modeling. \\

\section*{Entropy maximization: consistent inference in model-making}
In the inference-based perspective, the Maximum Entropy principle (MaxEnt) is the unique procedure to generate probability distributions that are logically consistent and match known externally supplied information.  In this view, probabilities are not limited to being frequencies of repeatable events -- they can characterize uncertainties of all kinds.  This is the Bayesian school of thought, in which we can discuss the probability of a nuclear disaster, or life on Mars, or the chance of snow at some particular time and date in the future, even though none of these are repeatable events.  Beginning with the 1957 Physical Review paper by E.T. Jaynes\cite{jaynes_information_1957}, the foundations of statistical physics have been re-framed in terms of Bayesian inference in order to avoid the need for replicates of physical systems.  \\

Jaynes saw statistical physics as a procedure for inference and model-making rather than as a law of nature.  He advised asserting only the information that you have (for example, average energy $\langle \varepsilon \rangle$ in the textbook\cite{huang_statistical_1987, dill_molecular_2010} case described above), and asserting maximal ignorance otherwise.  Jaynes' arguments removed the dependence on replicability of the event, but some physicists objected that statistical physics should not be tied to an observer's state of knowledge\cite{denbigh1985entropy}.  \\

This conceptual problem was resolved in the seminal work of Shore and Johnson in 1980\cite{shore_axiomatic_1980}, in which they fully axiomatized the inference process.  They showed that, given a \textit{prior distribution} $q_m$ representing some initial information about the system, and given new information constraining the possible forms for the \textit{posterior distribution} $p_m$, there is only one posterior distribution $p_m^*$ obeying consistency with the logical structure of probabilities which introduces no unwarranted bias; the unique posterior $p_m^*$ is that which maximizes the relative BG entropy of the posterior with respect to the prior, constrained by the relevant information (alternatively, minimizes the Kullback-Leibler (KL) divergence $D(p||q)$ of the posterior with respect to the prior):
\begin{equation}
S_{\mathrm{BG}} = - \sum\limits_m p_m \ln \left(\frac{p_m}{q_m} \right) = -D(p||q).
\label{eq:fqs}
\end{equation}
Although $S_\mathrm{BG}$ can be computed for all possible posteriors with respect to the given prior, it is only the maximum value $S_\mathrm{BG}^*$ that selects the proper posterior $p_m^*$.  Up to a constant, this relative $S_\mathrm{BG}$ reduces to those described above in Eqns. \eqref{eq:fs}-\eqref{eq:maxent} when the prior is uniform; the common assumption of uniform prior, known as the \textit{principle of equal a priori probabilities}, is sometimes justified by symmetry arguments.  For example, if the microstate refers to the position of a particle, and the system has isotropic symmetry, the starting point of knowledge is that no position should be different than any other position, so all positions are equally likely.  \\

\begin{figure}
    \centering
    \includegraphics[width=0.85\columnwidth]{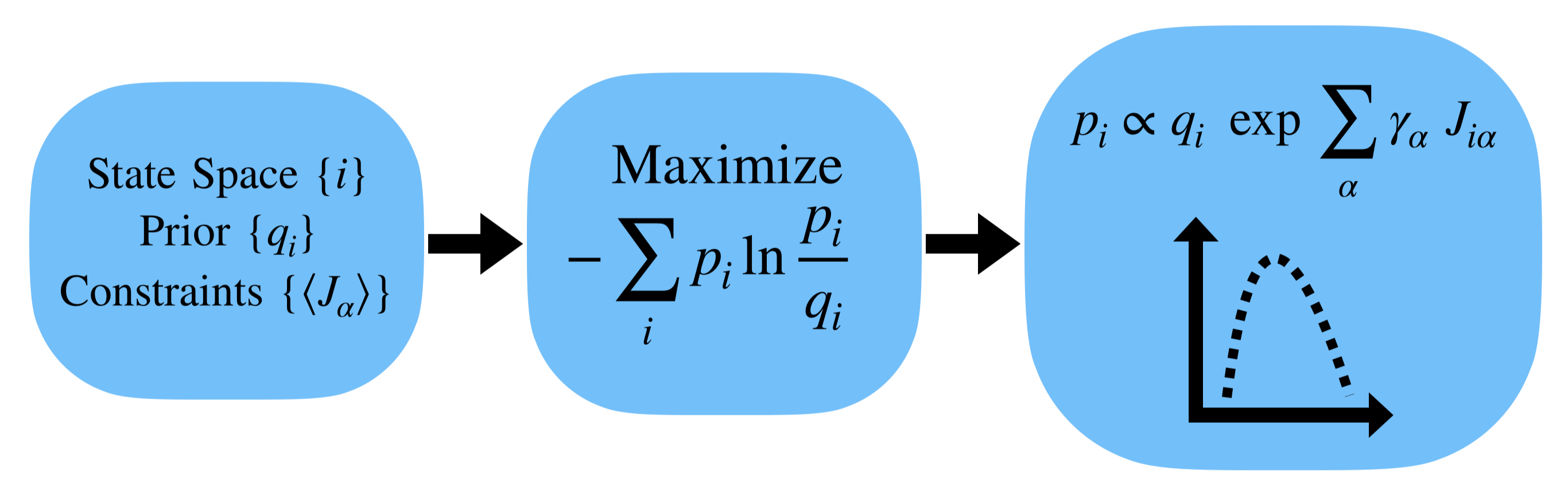}
    \caption{\textbf{The recipe of statistical physics:} A model-maker provides the ``physics'': the relevant state space, prior distribution, and constraints on possible posteriors; these serve as inputs for the Maximum Entropy (MaxEnt) inferential procedure -- ``the statistics'', which generates the best possible probability distribution given those inputs}
    \label{fig:pipeline}
\end{figure}

In this inference-based perspective (see Fig. \ref{fig:pipeline}), statistical physics entails: (i) a user-defined state space, (ii) a user-defined prior distribution, (iii) appropriate user-defined constraints, such as known averages from external data, and (iv) the MaxEnt procedure, predicting the posterior distribution that faithfully represents the premises and priors without unwarranted bias.  In this view, Maximum Entropy is not a law of physics; rather, Maximum Entropy is a method of drawing inferences. A user asserts a base \textit{model} structure (i.e. the state space, prior, and constraints -- the input ingredients needed for the MaxEnt recipe) and computes a posterior distribution.  If a predicted distribution is found to be inconsistent with experiments, the burden is on the user to find a better model. \\

\paragraph{``Non-extensive'' statistical physics.}  MaxEnt typically yields exponential distributions.  However, many systems, ubiquitous throughout nature, display non-exponential distributions, such as power-law tails\cite{bak_how_1996}.  What approach would predict non-exponential distributions?  A field called “non-extensive” statistical physics \cite{10.1093/oso/9780195159769.003.0006} derives non-exponential distributions using non-BG entropies, such as Tsallis or Renyi entropies.  However, non-BG entropies 
introduce unwarranted bias\cite{presse_nonadditive_2013,tsallis_conceptual_2015,presse_reply_2015}.  
The key axiom in Shore and Johnson that prevents that is what Jizba and Korbel called {strong-system independence}\cite{jizba_maximum_2019} and what Caticha summarized as the combination of {subset independence} and {minimal updating}\cite{caticha_entropy_2021}: \textit{When two systems are a priori assumed to be independent and new information does not reveal whether the systems are independent or not, the independence in the prior should be maintained.} Proper inference without unwarranted bias can only be done with the BG entropy and its maximization.  \\

How can non-exponential distributions be explained on the basis of the BG entropy?  This readily occurs for certain types of correlations in the imposed constraints.  For example, in social and financial situations, an ``economy of scale'' model gives non-exponential distributions with power-law tails \cite{doi:10.1073/pnas.1320578110}.
Exponentials arise when the function being constrained, which in classic statistical physics systems is the energy $\varepsilon_m$ is any function $f(m)$ of only the state index $m$.  However, if the energy or energy-like function $\varepsilon_m = g(m, \varepsilon_n)$ is a function both of state index $j$, and also of the energy values of other states\cite{doi:10.1073/pnas.1320578110}, power-law tails emerge naturally from the maximization of BG entropy.  Other methods yielding non-exponentials from BG entropy include superstatistics (which itself is simply a nested MaxEnt procedure) \cite{davis_conditional_2020, davis_derivation_2023, davis_q-canonical_2022, van_der_straeten_superstatistical_2008}, non-ideal reservoirs \cite{ramshaw_maximum_2022, vasconcelos_maximum_2018, pachter_nonequilibrium_2023}, and others \cite{umpierrez_fluctuation_2021, bercher_tsallis_2008, hernando_variational_2012}. \\

\section*{Beyond equilibria: defining \textit{irreversibility}}

\begin{figure}
    \centering
    \includegraphics[width=\columnwidth]{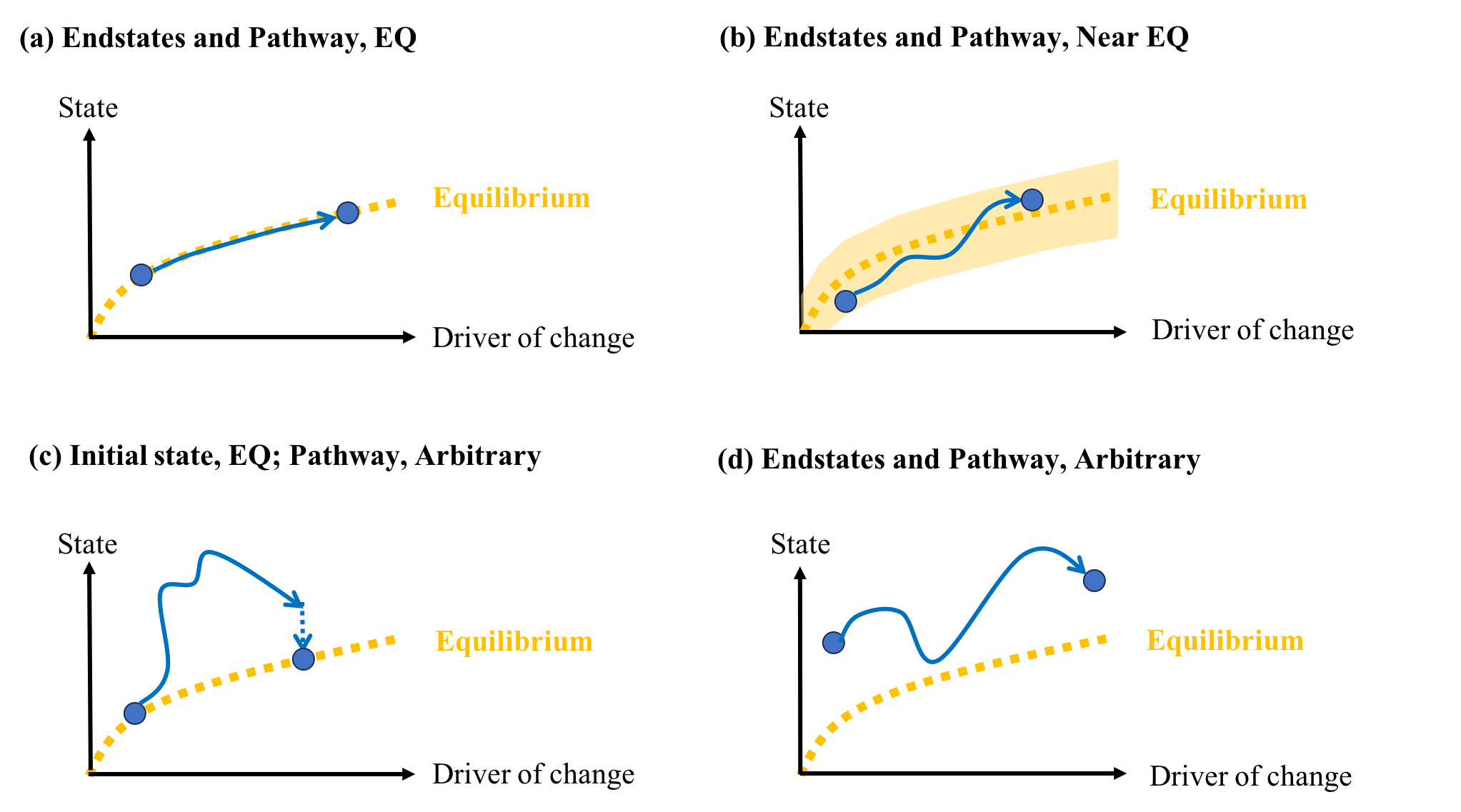}
    \caption{
    \textbf{Four classes of dynamical change:} ``State'' here in the vertical axis means the macroscopic/thermodynamic state of the system.  (a) Starting from one equilibrium, ending at another equilibrium through slow reversible processes. Clausius' equality applies. (b) From one near-equilibrium to another near-equilibrium through an arbitrary near-equilibrium path, the subject of stochastic thermodynamics. Shaded area indicates near-equilibrium of a system embedded in equilibrium baths. (c)  From an initial equilibrium to an arbitrary end state through an arbitrary non-equilibrium path, the subject of Clausius inequality and Jarzynski equality. The generally non-equilibrium end state could be related to an equilibrium state through a relaxation process where no work is performed (indicated by the dashed-line arrow), a special case that is the subject of Boltzmann's H Theorem.   (d) Arbitrary non-equilibrium process, the subject of Maximum Caliber.}
    \label{fig:5themes}
\end{figure}

We now turn to \textit{non-equilibrium} statistical physics.  Fig \ref{fig:5themes} divides dynamical processes into four classes of behaviors.  Described in more detail below, they are briefly:  (Fig \ref{fig:5themes}a) Largely as described in the Clausius entropy section above -- begin in initial equilibrium (EQ) state and end in final EQ state, proceeding through an idealized extremely-slow, quasi-equilibrium pathway.  (Fig \ref{fig:5themes}b) Begin in an arbitrary near EQ state and end in an arbitrary near EQ state through an arbitrary near EQ path -- this is the subject of Stochastic Thermodynamics (ST) and Macroscopic fluctuation theory; see below.  (Fig \ref{fig:5themes}c) Begin in initial EQ state, take an arbitrary path to an arbitrary point, then relax to a final EQ state under fixed external conditions -- this is the purview of the Clausius inequality and Jarzynski relation\cite{jarzynski_equalities_2011}.  The final relaxation segment is the realm of the Boltzmann's H Theorem.  (Fig \ref{fig:5themes}d) Begin in an arbitrary state and end in arbitrary state with arbitrary connecting pathway. This is the most general class, and it is the realm of the principle of Maximum Caliber. \\

We note that ``near-equilibrium'' in the classification above (and in our later discussions) means something different from what one might expect.  Consider a total super-system, consisting of a system of interest that can hypothetically be arbitrarily far from EQ, coupled to bath(s); if the bath(s) remain perpetually in EQ, then we refer to the super-system as ``near EQ'', because a system that is driven strongly enough will eventually perturb its environment noticeably out of EQ\cite{pachter_nonequilibrium_2023}.  Note that this is \textit{NOT} the same as what ``near-equilibrium'' means in stochastic thermodynamics (ST), where the term is reserved for processes in which the state of the system is always close to the equilibrium distribution. In ST, when the system is said to be near-equilibrium, linear response theory applies, whereas linear response theory breaks down when the system is far from equilibrium in ST.\\

A central question:  Can the various forms of irreversibility in nature be quantified by some entropy-like variational principle?  We review below successes in modeling in the four classes of processes above.  We note that local-equilibrium -- \emph{a.k.a.} local-detailed-balance -- approximations, and the assumption of infinite equilibrium heat baths, can limit the generality of most of the methods.  Maximum Caliber, however -- a variational principle that maximizes the \textit{path entropy} -- is free from such limitations. \\

\paragraph{Boltzmann's H Theorem and Irreversible Relaxation}  In seeking connections between microscopic particles and macroscopic thermodynamics, Boltzmann considered the following entropy-like functional of the distribution density $f(x,v,t)$ for particles in a gas to have position $x$ and velocity $v$ at time $t$:
\begin{equation}
H = - \int \int \mathrm{d}v \mathrm{d}x \ f(x,v,t) \ln f(x,v,t)
\end{equation}
which reduces to the classic Boltzmann entropy when all microstates are equally probable. \\

Boltzmann used this quantity to state his $H$-theorem: namely, that $H$ increases in time until the gas equilibrates, at which point $H$ attains its maximum value equal to the equilibrium entropy; see Fig  (\ref{fig:5themes}c), and see Huang's textbook\cite{huang_statistical_1987} for more detailed information.  Conceptually, this is an early example of what mathematicians now call a \emph{Lyapunov function} -- a quantity whose changes in time are monotonic for a particular dynamics and thus characterize the irreversible relaxation to steady state.  \\

For stochastic dynamics or dissipative deterministic dynamics, Lyapunov functions have proven useful for, among other things, characterizing the direction of time. For Markov dynamics with a steady-state, a well-known Lyapunov function is the relative entropy\cite{cover_elements_2006}, i.e. the KL divergence
\begin{equation} \label{KL divergence}
    D(t) = D(p(t)||\pi)=\sum_i p_i(t) \log \frac{p_i(t)}{\pi_i}
\end{equation}
where $p_i(t)$ is the probability distribution at time $t$ and $\pi_i$ is the steady-state probability distribution.  For Markov dynamics with an unique steady state, $\mathrm{d}D(t)/\mathrm{d}t\le 0$, which characterizes the stability of the steady state\cite{schnakenberg_network_1976} and also bounds the maximum extractable work in a transition between arbitrary distributions\cite{parrondo_thermodynamics_2015}.  Studies of the geometrical origin of these results grew into the field of \emph{information geometry}\cite{amari_information_2016}, which continues producing new insights, such as an upper limit on the relaxation speed of an observable related to its fluctuations and the local curvature of $D(t)$\cite{crooks_measuring_2007,nicholson_time-information_2020,ito_stochastic_2020}. \\

For dissipative deterministic dynamics, a Lyapunov function known as the ``\emph{quasi potential}'' can be derived by considering the deterministic dynamics as the zero-noise limit of a stochastic dynamics\cite{freidlin_random_1984}.  This Lyapunov function is useful in analyzing limit cycles\cite{strogatz_nonlinear_2018}, Kramers' theory of barrier-crossing in chemical reaction kinetics\cite{freidlin_random_1984,gardiner_stochastic_2009}, and nonequilibrium dynamics in living systems in general\cite{nolting_balls_2015,fang_nonequilibrium_2019,agozzino_how_2020}. More recently, the quasi potential has been given a stochastic thermodynamics interpretation\cite{qian_kinematic_2020,yang_potentials_2021}, making it conceptually unified with the stochastic Lyapunov function described above.   \\

\paragraph{Stochastic Thermodynamics (ST) computes properties of irreversibility.}  The field of \textit{Stochastic Thermodynamics} (ST)\cite{jarzynski_equalities_2011,seifert_stochastic_2019,peliti_stochastic_2021} -- which inherits key concepts from chemical thermodynamics\cite{prigogine1962chemical} and from Schnakenberg's chemical-reaction network physics\cite{schnakenberg_network_1976} -- has emerged in the last few decades, generalizing Clausius' theorem in order to more broadly quantify irreversibility; see Fig  (\ref{fig:5themes}b) (and (\ref{fig:5themes}c) for Jarzynski's equality of Hamiltonian systems\cite{jarzynski_hamiltonian_2000,jarzynski_equalities_2011}). \\

Note that Eq. \eqref{delta S_C lower bounds heat dissipation} from Clausius admits a separation for any process starting and ending in equilibrium into two components, 
\begin{equation} \label{heat-decomp}
\int_{\mathrm{general}} \frac{\delta Q}{T}  = -\Delta S_{\mathrm{Clausius}} +I_{\mathrm{general}}
\end{equation}
where $-\Delta S_{\mathrm{Clausius}}$ is the heat dissipation integral computed along a reversible trajectory.  So the total heat dissipation integral of an arbitrary process can be parsed into two contributions: the baseline dissipation one would get for a reversible process, and the remaining term $I_{\mathrm{general}} \geq 0$ measuring the extra heat dissipation due to irreversibility, i.e. it quantifies the degree of deviation from reversibility.  \\

The original Clausius picture is limited in two ways: (1) The starting and ending states must be equilibria, so it cannot apply to arbitrary dynamical processes, and (2) a well-defined temperature must exist at all points along the transformation, which is equivalent to assuming an idealized thermal bath.  The goal of ST is to analyze the irreversibility of general processes beyond these limitations (although the second limitation still applies to the prominent interpretation of many ST results).\\

At the heart of ST is the quantity $\delta A_{ij}$ called \emph{affinity}, associated with transitions between states $i$ and $j$.  In Markov processes described by transition rates $k_{ij}$, the \emph{affinity} is defined as 
\begin{equation} \label{affinity def}
    \delta A_{ij} = \log \left(\frac{k_{ij}}{k_{ji}} \right).
\end{equation}
The necessary and sufficient condition for a Markov process to be detailed-balanced is for the affinity to be zero for any cycles.  This is the Markov process analogue of Clausius' theorem, and is also known as the Kolmogorov criterion in mathematics\cite{jiang_mathematical_2004,yang_bivectorial_2021}.  In further analogy with the Clausius picture, the affinity can be decomposed into an equilibrium-like landscape-descending term and an additional term that quantifies the breakdown of detailed balance,
\begin{equation} \label{affinity decomposition}
    \delta A_{ij} = \log \left(\frac{k_{ij}}{k_{ji}}\right) = -\Delta (-\log \pi) +\log \left(\frac{\pi_i k_{ij}}{\pi_j k_{ji}} \right)
\end{equation}
where $\pi_i$ is the steady-state distribution. The steady-state average of Eq. \eqref{affinity decomposition}, denoted by angle brackets $\langle \cdot \rangle$, gives the Markov process generalization of Eq. \eqref{heat-decomp}:
\begin{equation} \label{average affinity decomposition}
    \langle \delta A \rangle = \sum_{i<j} J_{ij} \log \left(\frac{k_{ij}}{k_{ji}}\right) = -\Delta \langle -\log \pi \rangle + \underset{\ge 0}{\underbrace{\sum_{i<j} J_{ij}\log \left( \frac{\pi_i k_{ij}}{\pi_j k_{ji}} \right)}}
\end{equation}
where $J_{ij} = \pi_i k_{ij} - \pi_j k_{ji}$ is the steady-state probability flux.\\ 

The affinity decomposition of Eq. \eqref{affinity decomposition} exemplifies the key idea of ST, inherited from Clausius: dynamical properties can be split into a detailed-balance component and an additional component related to breaking of detailed-balance.  This type of decomposition appears in many places, with arguably the best-known result\cite {ge_extended_2009,esposito_three_2010} parsing the total irreversibility in nonequilibrium Markov processes $I_{\mathrm{tot}}(t)$ into two contributions: (i) the deviation from its steady state (non-stationarity), quantified by $-\mathrm{d}D(t)/\mathrm{d}t\ge 0$ as defined in Eq. \eqref{KL divergence}, and (ii) the additional irreversibility due to detailed-balance-breakdown
\begin{equation}
    I_{\mathrm{DBB}}(t)=\sum_{i,j} p_i(t) k_{ij} \log \frac{\pi_i k_{ij}}{\pi_j k_{ij}} \ge 0
\end{equation}
The total irreversibility rate in ST is the KL divergence of the probability of a two-state trajectory with respect to its time reversal:
\begin{equation} \label{total irreversibility decomposition}
    I_{\mathrm{tot}}(t) = \sum_{i,j} p_i(t) k_{ij} \log \frac{p_i(t)k_{ij}}{p_j(t) k_{ji}}= -\frac{\mathrm{d}D}{\mathrm{d}t}(t) + I_{\mathrm{DBB}}(t)
\end{equation}
For detailed-balanced systems, irreversibility comes solely from the relaxation to the steady state, characterized by the monotonic behavior of the first term $-\mathrm{d}D(t)/\mathrm{d}t$.  If, however, the system has broken detailed-balance, it has non-zero probability flux even at steady-state, which produces a non-zero $I_{\mathrm{DBB}}$; in other words, this second term captures the irreversibility due to an underlying non-equilibrium steady-state.  \\

Not only the total irreversibility, but also the two components -- the non-stationarity and the detailed-balance-breakdown -- can be expressed as KL divergences of path probabilities for forward processes with respect to the probabilities of reversed processes\cite{crooks_path-ensemble_2000,ge_extended_2009,esposito_three_2010,yang_unified_2020}.  These path-probabilistic expressions, which can be considered as the mathematical definitions of these notions of time irreversibility, lead to a zoo of fluctuation relations and uncertainty relations for ST of Markov processes\cite{crooks_path-ensemble_2000,jarzynski_equalities_2011,yang_unified_2020} and help identify the domain of validity for the various forms of fluctuation relations\cite{yang_unified_2020}.  \\

The same decomposition technique proves useful in analyzing other quantities.  For example, a generalized Einstein relation exists\cite{yang_time-translational_2023} in which the detailed-balanced component of the covariance between the system's state and its rate-of-change is related to fluctuations.  Another example is the velocity and flux decomposition for diffusion processes  \cite{graham_covariant_1985,ao_potential_2004,wang_potential_2008,yang_potentials_2021} and its zero-noise limit\cite{freidlin_random_1984,yang_potentials_2021}.  Velocity and flux in these systems are decomposed into gradient-descending parts (the detailed-balance component) and a complementary cyclic part embodying the breaking of detailed-balance.  This is actually the previously mentioned method for finding a Lyapunov landscape function\cite{freidlin_random_1984,qian_kinematic_2020,yang_potentials_2021}, which exemplifies the fundamental insufficiency of describing a non-detailed-balanced system solely by its scalar landscape function. \\

\paragraph{The Local Detailed Balance Assumption in Stochastic Thermodynamics}
A key ingredient that ST uses to provide physical meaning to mathematical results is a condition known as Local Detailed balance (LDB); this is an assumption that all transitions of a system occur through exchanges with perfect thermodynamic reservoirs, which are infinite in size and always in equilibrium (though there exists some work in ST regarding NEQ baths\cite{Cockrell_Ford_2022}).  The system may be in contact with multiple distinct reservoirs\cite{esposito_stochastic_2012}, 
but exchanges with each reservoir are assumed to occur independently of each other, so that each transition of the system can be associated with an exchange of conserved quantity with exactly one bath.  Then, for example, we can equate the affinity $\delta A_{ij}$ for heat bath-induced transitions with the physical heat dissipation divided by temperature $\delta Q/T$, where $T$ is the temperature of the heat bath associated with that particular transition, and similarly for exchanges of other quantities\cite{sekimoto_stochastic_2010,seifert_stochastic_2018}. \\

On one hand, LDB only enters when we wish to understand affinity in terms of heat and temperature\cite{yang_bivectorial_2021} or other physical quantities. Many mathematical results in ST, such as Eq. \eqref{affinity decomposition}, speed limit relations\cite{nicholson_time-information_2020,ito_stochastic_2020}, fluctuation relations\cite{crooks_path-ensemble_2000,jarzynski_equalities_2011,yang_unified_2020}, uncertainty relations\cite{barato_thermodynamic_2015,horowitz_thermodynamic_2020}, and others remain valid and useful for the analysis of Markov stochastic dynamics in general, even if LDB does not hold.  On the other hand, we must be aware of the limitations of applicability of LDB: we\cite{pachter_nonequilibrium_2023} have previously discussed the limits of the perfect bath assumption and how it essentially constrains the system to not be driven too strongly away from EQ; Falasco and Esposito \cite{falasco_local_2021} show that LDB holds only in the limit of weak driving and/or weak noise; Hartich and Godec \cite{hartich_violation_2022} show that LDB might not hold in a coarse-grained system even when there is clear time-scale separation; and Wasnik \cite{wasnik_revisiting_2023} disproves the generality of the decomposition of transition rates due to multiple reservoirs into the sum of transition rates due to each reservoir. \\ 
\begin{figure}
    \centering
    \includegraphics[width=\columnwidth]{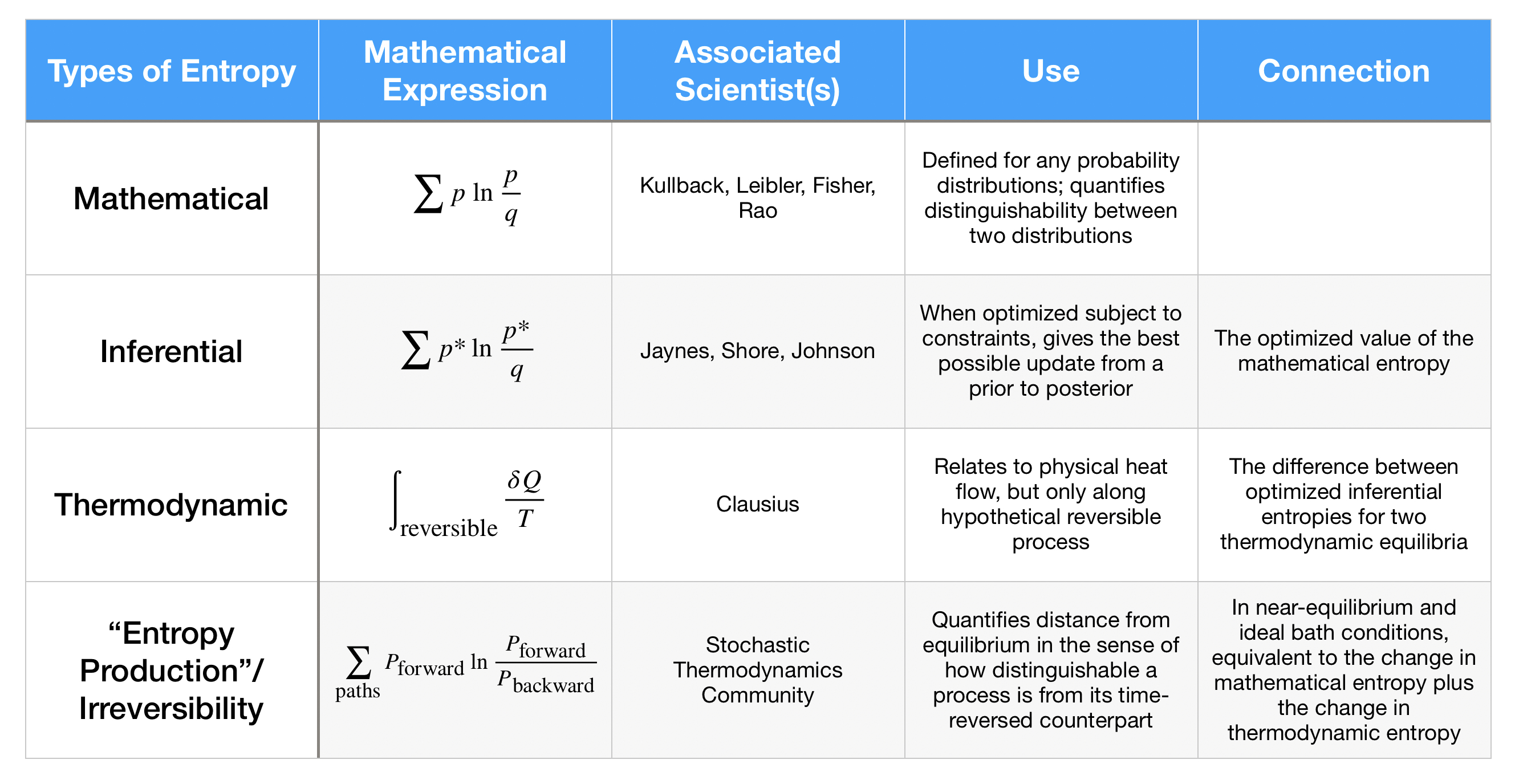}
    \caption{\textbf{Different types of \textit{entropy}, their uses, and the connections between them}}
    \label{fig:table}
\end{figure}

When LDB holds, the irreversibility $I_{\mathrm{tot}}(t)$ can be written as the change in BG entropy of the system plus the change in Clausius entropy of the environment, which lead the ST community to call the total irreversibility ``Entropy Production''.  However, this terminology can lead to confusion, since it conflates thermodynamic and inferential entropy, which are only equivalent for quasi-static, reversible processes, as we described in a previous section.  Furthermore, inferential entropy is not a tangible quantity that can be produced during a process, with a measurable difference in quantity from the beginning to the end.  Since the usefulness of $I_\mathrm{tot}$ goes beyond situations where LDB is applicable, we recommend exclusively using the term ``irreversibility''.  See Fig. \ref{fig:table} for further clarification of the distinct notions of entropy and their uses.  \\

\paragraph{Jarzynski Equality}
After work by Hunter and Reinhardt\cite{reinhardt_variational_1992,hunter_finitetime_1993} relating to computational methods, Jarzynski independently showed how to generalize Clausius' inequality of work $W$ and free energy $F$ into an equality\cite{jarzynski_nonequilibrium_1997}: 
\begin{equation}
    \langle e^{-W/k_B T} \rangle = e^{-\Delta F/k_B T}.
\end{equation}
There are technically two versions of the Jarzynski equality (JE): the original, as derived in 1997, applies to Hamiltonian dynamics for a system-plus-bath super-system that starts in equilibrium and gets driven arbitrarily away from EQ\cite{jarzynski_equalities_2011} before relaxing back to equilibrium once the external driving is completed -- this is depicted in Fig. (\ref{fig:5themes}c).  This Hamiltonian formulation of the JE does not need the LDB assumptions (although it does need another external perfect heat bath described by temperature $T$ with which the total super-system can equilibrate), and it can be generalized to strongly-coupled systems\cite{seifert_first_2016,jarzynski_stochastic_2017,miller_entropy_2017}.\\

For Markov dynamics as typically employed in ST, an equality of the same form as the original JE can be derived\cite{crooks_nonequilibrium_1998}. As with many other results in ST, while the mathematical statement is general\cite{yang_unified_2020}, the understanding of Markov transition probabilities in terms of free energy change and work requires the LDB assumption.  Hence, as with ST, the JE of Markov dynamics belongs in the classification of Fig. (\ref{fig:5themes}b), where the typical physical interpretations are only valid for systems embedded in idealized equilibrium baths.  Similarly, for other fluctuation relations\cite{spinney2013fluctuation}, the irreversibilities associated with them can lose their connections to heat dissipation when the bath itself is driven out of equilibrium.\\

\paragraph{Macroscopic Fluctuation Theory}
The field of Macroscopic Fluctuation Theory (MFT)\cite{bertini_macroscopic_2015} uses LDT to describe the non-trivial asymptotic behaviors and time irreversibility of stochastic systems in the hydrodynamic limit, in which the particles are so numerous that we can define continuous density and flow. This extends textbook hydrodynamics to the realm of fluctuations of density and flow in physical space-time. Since MFT also assumes a fast equilibrating bath for its physical interpretations, it faces the same LDB limitations as ST, and therefore can be put in to the same silo as ST in Fig.  (\ref{fig:5themes}b).  Both ST and MFT bring us powerful mathematical tools for analyzing stochastic processes and their irreversibilities, with results framed in terms of entropy-like quantities that provide direct physical insight in certain circumstances. Yet, a \textit{generative} principle for building dynamical models based on prior assumptions and data -- like what entropic inference has done for equilibrium physics and beyond -- seems to be missing in these approaches to non-equilibrium physics. \\

\section*{Towards a generative statistical physics of non-equilibria}

Equilibrium statistical physics has stood alone on a pedestal for more than a century.  The rigorous foundations of equilibrium statistical physics, combined with its breadth of applicability and generative power have -- until recently -- been unequalled by any treatment for non-equilibrium.  This was well articulated by Touchette\cite{touchette_large_2009}:  
\begin{quote}
    \textit{There is no general principle whereby one can calculate the distribution of the system's states from the sole knowledge of the system's invariants or external constraints imposed on the system … Such a general principle is precisely what a statistical-mechanical ensemble is, and what is missing from the theory of nonequilibrium systems.}
\end{quote}

Said differently, Touchette's aspiration is for a \textit{generative principle.} Yes, NEQ physics has specific models -- of random flights, Langevin and Fokker-Planck diffusion, and others, and useful mathematical results for their analysis, many in ST, LDT, and MFT. But, as indicated by Fig~\ref{fig:5themes}, NEQ modeling has largely been limited to starting and ending points of processes that are at or near equilibrium and they require contact with a perfect bath.  Such theories do not \textit{derive} their probability distribution functions (i.e. rates and routes for dynamics) from a model and data, as MaxEnt does for EQ.  Rather these approaches \textit{assume} distributions, and in particular, they assume \textit{equilibrium distributions.}  Touchette's aspiration is for some enveloping principle for making models that can derive NEQ dynamical distribution from the model and data.  We argue below that: (i) the problems referred to in Touchette's remark largely resulted from efforts that were too closely tied to EQ principles, and (ii) the principle of Maximum Caliber appears to fill this gap.  \\

\paragraph{\textit{Entropy production rate} is not the basis for NEQ variational principle.} 
Historically, there were efforts to seek a NEQ variational principle in the time rate of change of the entropy of the system state $\mathrm{d}S_{\mathrm{BG}}(\{p_i(t)\})/\mathrm{d}t$ or the entropy production of the total super system -- either its minimization\cite{PhysRev.96.250,de_groot_non-equilibrium_2011}, or maximization\cite{e11040931,Martyushev:2021}. While they can be useful in certain cases, there are counter examples, e.g. the Schl\"{o}gl model for bistable chemical system\cite{vellela_stochastic_2008}, demonstrating that neither of them is generally valid.
From the inference perspective, to infer $\{p_i(t)\}$ in a non-stationary system from time $0$ to $\tau$ requires applying MaxEnt to every instance $t\in[0,\tau]$. At each instance $t$,  $S_{\mathrm{BG}}$ is maximized by $p_i^*(t)$ among all $p_i$ satisfying the corresponding constraints, but this says nothing about the derivative of $S_{\mathrm{BG}}(\{p_i^*(t)\})$ being maximized or minimized. Furthermore, the above procedure is not enough for a NEQ variation principle of dynamics: there are more than one dynamics that can produce a given $\{p_i(t)\}$. This objective is met by the principle of Maximum Caliber, described below.  \\

\paragraph{Limitations of local detailed balance and heat baths.}

Assumptions of local detailed balance and infinite baths have been important and useful theoretical tools since the development of statistical physics, especially for the recent development of ST\cite{seifert_stochastic_2019,peliti_stochastic_2021}. Yet, it can be limiting to assume that fluctuations in a system correspond to an equilibrium reservoir, as commonly occurs in Langevin or Fokker-Planck models. 
It requires that a system's fluctuations are neither too large nor too fast, as described above. Rather, one requires a broader principle for generating non-equilibrium fluctuations that is not tied down by equilibrium ideas.  Furthermore, for broadest generality, a non-equilibrium principle should apply to force-flow relations that go beyond fluids, particles, electrical charges, and heat, to encompass also the flows of vehicles and goods along networks, of proteins and genes in cells, of electrical signals in brains, and of biological changes in organisms in evolution and ecology.   These requirements are fulfilled by the principle of Maximum Caliber, as we now describe.  \\

\section*{Maximum Caliber: the variational principle for non-equilibria}

A general and generative variational method for NEQ is the principle of Maximum Caliber.  Maximum Caliber (MaxCal) -- a concept and term originating with Jaynes\cite{jaynes_minimum_1980} -- does for trajectories what classic Maximum Entropy (MaxEnt) does for states.  In fact, it is identical to MaxEnt except that the probabilities involved refer to paths through state-space, rather than single states; it updates from a prior probability distribution to a unique posterior by incorporating any known constraints, satisfying the axioms of entropic inference that go back to Shore and Johnson. \\

Consider a prototypical example which follows almost exactly the form of the MaxEnt example given above in Eqs. \eqref{eq:maxent}-\eqref{eq:Z}.  We use Lagrange multipliers to maximize the Caliber $\mathcal{C}$, which is the BG entropy of paths, subject to constraints on normalization and some average flux quantity (similar to the average energy in the equilibrium example):
\begin{equation}
\mathcal{C} = -\sum_{k=\mathrm{paths}} p_k \log \left(\frac{p_k}{q_k}\right) - \gamma \left (\sum_k p_k J_k - \langle J \rangle \right ) + \alpha  \left ( \sum_k p_k - 1\right ).
\label{eq:caliber}
\end{equation}
In Eq. \eqref{eq:caliber},  $\gamma$ is a Lagrange multiplier that tunes the average $\langle J \rangle$ , while $\alpha$ ensures normalization.  The maximizing distribution is
\begin{eqnarray}
p_k^* = q_k \frac{e^{-\gamma J_k}}{Z_d}\label{eq:maxcal0}
\end{eqnarray}
where
\begin{eqnarray}
Z_d = \sum_k q_k e^{-\gamma J_k},
\label{eq:maxcal1}
\end{eqnarray}
a weighted sum over paths, is the dynamical equivalent of a partition function.  Thermodynamic-like force-flux relations between the conjugated variables of $\gamma$ and $\langle J \rangle$ can then be derived\cite{presse_principles_2013,yang_statistical_2023}.\\

Overviews of MaxCal have been given elsewhere \cite{presse_principles_2013, hazoglou_communication_2015, ghosh_maximum_2020, caticha_entropy_2021}; here is just a brief summary.  First, MaxCal recapitulates -- as it should -- many known results of near-equilibrium NEQ physics, including the Green-Kubo fluctuation-dissipation relations, Onsager reciprocal relations, Prigogine's minimum entropy-production principle in certain situations, and Kirchoff's current law.  MaxCal can derive the \textit{phenomenological laws} of force-flow relations, such as Newtonian viscosities of simple fluids, Fourier's Law of heat flow, Ohm's Law of electrical currents, and Fick's Law of particle flows.  It can also be shown to provide a basis for Markov modelling \cite{ge_markov_2012,lee_derivation_2012},
and it frames attempts to derive much of fundamental physics, such as Newtonian\cite{ge_analytical_2012,gonzalez_newtonian_2014} and quantum \cite{general_principle_2018,chetrite_e_2021} mechanics, from an entropic inference perspective. MaxCal -- and more broadly MaxEnt -- inference methods are employed successfully in many cutting edge areas like computational science\cite{Bolhuis_Brotzakis_Vendruscolo_2021, tsai_path_2022, ALVES20151} and theoretical endeavors like superstatistics\cite{sattin_bayesian_2006, caticha_entropy_2021}, as well as guiding procedures for dealing with systems with strong coupling \cite{dixit_maximum_2013} and non-ideal baths \cite{pachter_nonequilibrium_2023}, beginning to bridge the gaps outlined above regarding perfect bath assumptions, and therefore treating systems that are truly arbitrarily far from equilibrium.  MaxCal can successfully treat non-linearities, small systems, and it can be applied to all kinds of flows.  These successes plant the seeds for ever more promising MaxCal results.

\paragraph{How to choose constraints for the MaxCal procedure.}  As noted above, the burdens of modeling with MaxEnt or MaxCal lie with the user.  The user defines the model.  The user must also assert the appropriate constraints relevant to the process at hand.  To illustrate, first consider simply EQ modeling.  If you know only the temperature $T$ at a system's boundary, in a process where volume $V$ or particle number $N$ are also controlling, your predictions will be misinformed.  Similarly, modeling will err if it treats only bulk constraints $(T, V, N)$, where interfacial area $A$ also matters as an independent variable.  This is not a flaw of MaxEnt; it is the burden of the user to know which, and how many, variables matter for the problem at hand.  \\

The problem is identical for non-equilibrium modeling using MaxCal.  Consider a ball falling through a viscous liquid.  Suppose its average velocity is $\langle v \rangle$.  Now, measure the same ball dropping through a more viscous liquid, but at the same velocity $\langle v \rangle$ (hence driven by a larger force).  The ball's velocity alone is not sufficient to characterize this system -- the other relevant variable here is heat dissipation, $\langle q \rangle$; MaxCal modeling of microscopic rate distributions in this situation requires knowing both of these constraints\cite{PhysRevE.100.010105}.  Even in equilibrium modeling for a system in contact with a heat bath, if the system undergoes large changes, or fast changes, such that the heat bath cannot respond sufficiently, one must specify the thermal conductivity or the heat capacity of the bath\cite{pachter_nonequilibrium_2023}, in addition to its temperature.  For more discussion of constraints, see \cite{caticha_entropic_2013}.  \\

\section*{Conclusions}
The early conceptions of statistical physics saw systems as large ensembles of replicates, with probabilities seen as frequencies, and with assumptions of ergodicity and chaotic collisions.  A newer view holds that maximizing entropy is simply a way of drawing consistent inferences about probabilities, where the user is responsible for a proper model of physics, proper assumptions about the equivalencies among states, and proper choice of constraints. The same maximization of entropy procedure applies to non-equilibrium predictions of forces and flows, except one must use \textit{path entropies} instead of \textit{state entropies.} This opens up a broad area of dynamical modeling -- applicable beyond heat baths, beyond thermal materials, to situations far-from-equilibrium and with non-linearities, applicable even to few-particle distributions.  \\

\section*{Acknowledgements}
We are grateful to the Stony Brook Laufer Center for support.  We also thank Bill Cannon, Chris Jarzynski, Charles Kocher, Steve Presse, David Sivak, and Jin Wang for insightful comments and helpful feedback.  \\

\section*{Author contributions}
The authors contributed equally to all aspects of the article. 

\section*{Competing interests}
The authors declare no competing interests.

\bibliography{main}

\cleardoublepage

\end{document}